\documentclass[sigconf]{acmart}

\AtBeginDocument{%
  \providecommand\BibTeX{{%
    \normalfont B\kern-0.5em{\scshape i\kern-0.25em b}\kern-0.8em\TeX}}}

\setcopyright{acmcopyright}
\copyrightyear{2020}
\acmYear{2020}

\acmConference[]{Under review}{2020}{Earth}
\acmPrice{15.00}

\begin{document}

\title[VASG]{Visually Aware Skip-Gram for Image Based Recommendations}

\author{Parth Tiwari}
\affiliation{%
  \institution{IIT Kharagpur}
  \streetaddress{1 Th{\o}rv{\"a}ld Circle}
  \city{Kharagpur}
  \country{India}}
\email{parth.tiwari95@iitkgp.ac.in}

\author{Yash Jain}
\authornote{Both authors contribted equally to this work}
\affiliation{%
  \institution{IIT Kharagpur}
  \streetaddress{1 Th{\o}rv{\"a}ld Circle}
  \city{Kharagpur}
  \country{India}}
\email{yashjainjain1704@gmail.com}

\author{Shivansh Mundra}
\authornotemark[1]
\affiliation{%
  \institution{IIT Kharagpur}
  \streetaddress{1 Th{\o}rv{\"a}ld Circle}
  \city{Kharagpur}
  \country{India}}
\email{shivanshmundra1@gmail.com}

\author{Jenny Harding}
\affiliation{%
  \institution{Loughborough University}
  \city{Loughborough}
  \country{United Kingdom}
}

\author{Manoj Kumar Tiwari}
\affiliation{%
 \institution{National Institute of Industrial Engineering}
 \city{Mumbai}
 \country{India}}

\renewcommand{\shortauthors}{Tiwari, et al.}

\begin{abstract}
    The visual appearance of a product significantly influences purchase decisions on e-commerce websites. We propose a novel framework VASG (Visually Aware Skip-Gram) for learning user and product representations in a common latent space using product image features. Our model is an amalgamation of the Skip-Gram architecture and a deep neural network based Decoder. Here the Skip-Gram attempts to capture user preference by optimizing user-product co-occurrence in a Heterogeneous Information Network while the Decoder simultaneously learns a mapping to transform product image features to the Skip-Gram embedding space. This architecture is jointly optimized in an end-to-end, multitask fashion. The proposed framework enables us to make personalized recommendations for cold-start products which have no purchase history. Experiments conducted on large real world datasets show that the learned embeddings can generate effective recommendations using nearest neighbour searches. 
\end{abstract}



\maketitle

\section{Introduction}

Recommender Systems have been described as “an intuitive line of defense against consumer over-choice” \cite{10.1145/3285029}. With the rapid growth in the number of products available on e-commerce websites, this problem of "over-choice" is becoming more and more significant, especially in the case of clothing related products. In fact, fashion recommendation has attracted considerable attention in recent literature. Studies tackling this problem have shown improved performance by incorporating visual information into their recommendation procedure. Hence it is safe to conclude that the visual appearance of a product plays a significant role while making purchase decisions.

Image-based recommendation systems can be broadly classified into two groups - (i) Systems which incorporate images into a product rating/rank prediction function. These systems build upon well known Latent Factor models like Singular Value Decomposition (SVD) or Bayesian Personalized Ranking (BPR). \cite{10.1145/3331184.3331254, 10.5555/3015812.3015834, HeLinWanMcA16, 10.1145/2959100.2959152, 10.1145/3077136.3080797} (ii) Systems which learn latent representations (a.k.a embeddings) using image features and subsequently use embedding distance for making recommendations. These systems rely on extracting fine-grained embeddings which can capture user preference and/or product similarity \cite{10.1145/3240508.3240541, McATarShiHen15, 10.1145/3219819.3219890, DBLP:journals/corr/ShankarNAKC17, 10.1145/2964284.2967182}. Beyond images,  information coming from product metadata \cite{10.1145/2959100.2959160}, user-product information networks \cite{8355676}, review-text \cite{10.1145/3240508.3240541, 10.1145/3291060} etc. has also been leveraged for learning effective representations. 

Making recommendations in the cold-start setting is a key area where addition of visual information has shown significant performance improvement \cite{10.5555/3015812.3015834}. However, we observe that there is considerable variability in the definition of cold-start setting across literature. Some studies describe cold-start products as items that have fewer instances of positive feedback in the training set \cite{10.5555/3015812.3015834, HeLinWanMcA16, 8355676, 10.1145/3077136.3080797, 10.1145/3397271.3401252}, whereas \cite{10.1145/2959100.2959160} describes its cold-start scenario when certain product-pairs have zero co-occurrences while training. Recommending completely new products (i.e. products which are unseen during the training step) is a tougher problem which is studied less frequently \cite{10.1145/564376.564421, 10.1145/2645710.2645751, 5693971}. 

Systems falling under category (i) (as described above) rely on user and product latent factors coming from  interaction records, hence, they are incapable of handling new products unless product latent factors are estimated by other methods. Systems falling under category (ii) show better promise of handling this problem. Methods like \cite{DBLP:journals/corr/ShankarNAKC17, McATarShiHen15}, which make recommendations based on product-product similarity are capable of extracting latent representations for a new product. However, here user preference is ignored while feature extraction hence making personalized recommendations is not possible. Methods which attempt to capture preferences for individual users are better suited for handling this problem.

To this end, we propose a novel framework which jointly learns (i) user and product embeddings in a common latent space and (ii) a mapping function which transforms product image features to the same latent space. In specific, we augment the Skip-Gram model \cite{Mikolov:2013:DRW:2999792.2999959} with a decoder architecture where the decoder uses the product Skip-Gram embeddings to reconstruct their corresponding image features. Here Skip-Gram  maximizes the probability of co-occurrence of users and products in a Heterogeneous Information Network (HIN) while the decoder minimizes the image feature reconstruction loss. This decoder serves two purposes- (i) it incorporates visual information into the skip gram embeddings and (ii) it is later used for learning a mapping function which transforms product image features to the Skip-Gram embedding space. The learning objective of this architecture is dependent on the genre of input (users or products) observed while training. The Skip-Gram model is optimized in case of both users and products whereas the decoder is updated only for products in a multitask fashion. We term this model as VASG (Visually Aware Skip-Gram). 

The mapping from the product image feature space to VASG embedding space is subsequently learned and can be used for finding effective embeddings for cold-start products. VASG embeddings reflect user preference while simultaneously capturing a product's purchase history and visual appearance. Since users and products are represented in the same latent space, personalized recommendations can be made directly by searching for products in the user's vicinity. Our contributions can be summarized as -  

\begin{itemize}
    \item We propose a novel multitasking Skip-Gram architecture, VASG, which is trained on two different objectives depending on the genre of the input (users or products). This architecture enables us to learn a mapping which transforms product image features to the learned embedding space. 
    \item Extensive experiments are performed on real world datasets and VASG embeddings are compared to state-of-the-art recommendation systems. The performance for unseen cold start products is also studied. 
    \item We analyze the learned embeddings in detail to shed some light on the information captured by them. 
\end{itemize}

\begin{figure*}[h]
  \centering
  \includegraphics[width=\linewidth]{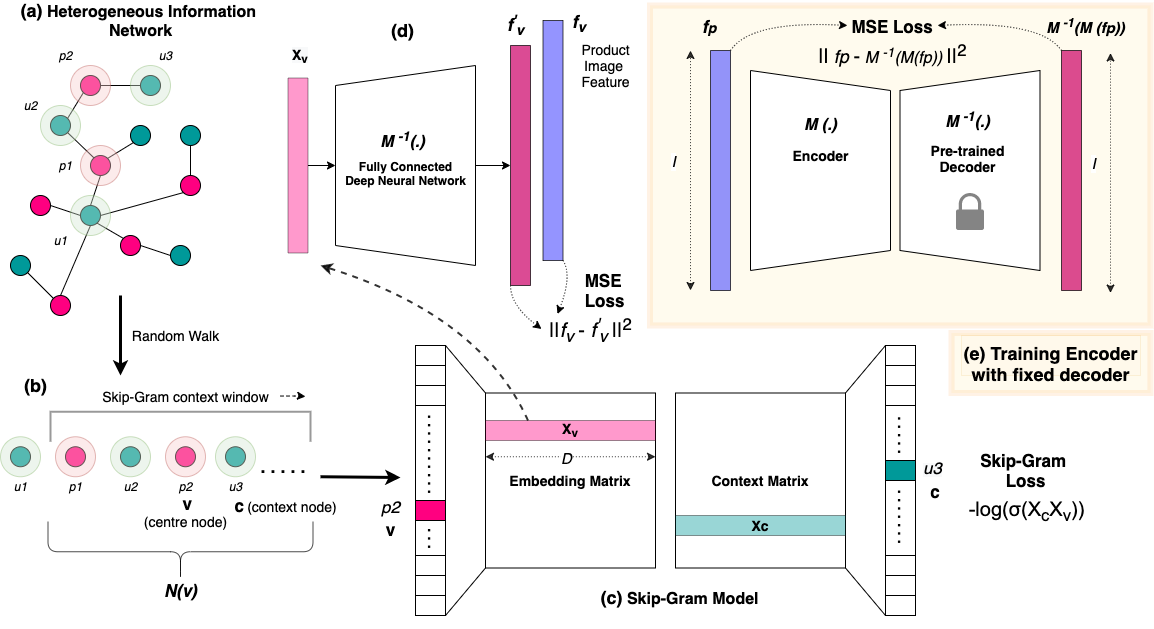}
  \caption{The VASG framework. (a) User-Product interactions are represented in a HIN. (b) Sequences are generated by performing random walks on the HIN. The Skip-Gram context window is used to describe the neighbourhood of a node. (c) The Skip-Gram model is trained on the generated sequences. (d) The decoder $\mathcal{M}^{-1}(.)$ attempts to reconstruct product image features from the Skip-Gram embeddings. (e) An autoencoder is trained to estimate the mapping $\mathcal{M}(.)$}
\end{figure*}

\section{Related Work}

Our work relies on several advancements in image feature extraction methods, representation learning techniques in graphs and multitask learning.

With advancements in deep learning, VBPR \cite{10.5555/3015812.3015834} established that the addition of visual signals can be useful for making recommendations. VBPR adds image features extracted from a pre-trained convolutional neural network to the BPR framework. Similarly, \cite{10.1145/3077136.3080797} extracts more fine grained image features using attention modules and incorporates them in collaborative filtering. Images have also been proven to be useful for Point-of-Interest recommendations in \cite{wang2017your} and for tag recommendations in \cite{rawat2016contagnet}. 

More recently, there has been a surge of methods which combine information coming from different modalities and metadata with images \cite{10.1145/3291060, HeLinWanMcA16, 10.1145/3240508.3240541, 10.1145/2959100.2959152, 10.1145/3331184.3331254}. Attention mechanisms which utilize textual reviews have been used for finding specific parts of an image where users are most interested in \cite{10.1145/3331184.3331254}. Reviews and image features have also been used for learning user preference which is then integrated into a rating-based matrix factorization model \cite{10.1145/3291060}. Pre-defined product categories and hierarchy trees have been leveraged for learning hierarchical product embeddings \cite{HeLinWanMcA16}.

Recommendation Systems which rely on product image similarity; for example methods which address the Street-to-shop recommendation problem \cite{6248071}; have received a tremendous boost with improvements in metric learning methods \cite{8575408, 10.1145/2964284.2967182, DBLP:journals/corr/ShankarNAKC17} . These methods use triplet loss to learn embeddings from product images which are robust to change in background, pose, lighting conditions etc. and hence can be used for making cross-domain recommendations. However, learning both user and product representations in a joint latent space is less frequently explored in existing literature. \cite{10.1145/3336191.3371770} embeds users and topics to the same low dimensional space to capture of their mutual dependency while \cite{10.1145/3240508.3240541} attempts to bring users, products and product search queries to the same latent space using both textual and visual modalities.

Heterogeneous Information Networks (HINs) can naturally model complex user and product interactions. Traditionally, meta-path based similarity and link predictions have been used for making recommendations in networks. With the development of graph representation learning methods \cite{perozzi2014deepwalk, grover2016node2vec, dong2017metapath2vec}, network embedding based recommendation systems have shown improved performance. The study \cite{8355676} learns user and product representations from a HIN and incorporates them into the {\it SVDFeature} \cite{Chen:2012:STF:2503308.2503357} framework. This idea of learning network representations has also been extended by using multitask learning for jointly optimizing the tasks of recommendation and link prediction in a HIN \cite{9051843}. Multitask learning is a training regime where multiple objectives are optimized simultaneously. It has proven to provide performance boosts in works like \cite{Gidaris_2019_ICCV, 8771379}.

A few studies which are relevant to our proposed approach should be discussed here.
Learning attribute-to-feature mappings has been proposed in \cite{5693971}. Here k-nearest neighbours, least squares approximation is used for approximating latent factors for new users or products. 
A large scale recommender system which uses images along with the HIN structure is proposed in \cite{10.1145/3219819.3219890}. Here highly efficient Graph-Convolutional Neural Networks are used on random walks generated from the HIN and recommendations are then made using similarity between the learned embeddings. 
Contrastive Predictive Coding \cite{oord2018representation} (CPC) is a generic representation learning regime which uses a similar training procedure as the proposed framework. CPC uses encoders to map raw data to latent representations and subsequently trains auto-regressive models using a contrastive loss. VASG differs from CPC in terms of usage of a decoding architecture instead of an encoder; usage of Skip-Gram instead of an auto-regressive model; and training using auxiliary supervision instead of self-supervision.

To the best of our knowledge, the model architecture proposed in this paper has not been studied before. In addition, we also make a novel attempt to bring unseen cold start products to the same latent space as existing users and products using product image features.

\section{VASG Framework}

Consider a set of users $\mathcal{U}$ and a set of products $\mathcal{P}$ where each user $u\in\mathcal{U}$ has interacted with a subset of products $\mathcal{P}\textsubscript{\it u} \subset \mathcal{P}$. Each product $p\in\mathcal{P}$ is associated with a visual feature $f\textsubscript{p}\in\mathbb{R\textsuperscript{I}}$ which is extracted from its image using a Deep Convolutional Neural Network. Only a subset of products $\mathcal{P}\textsubscript{warm} \subset \mathcal{P}$ is observed during training and is termed as the {\it warm start} set. The set of products which is never observed during training is termed as {\it completely cold} set $~\mathcal{P}\textsubscript{cold} \subset \mathcal{P}$ . It is to be noted that $\mathcal{P}\textsubscript{cold} \cap \mathcal{P}\textsubscript{warm} = \phi$.\\ Our objective is to learn (i) a {\it D} dimensional latent representation  $\mathcal{X}\in\mathbb{R}\textsuperscript{\it D}$ for each $u\in\mathcal{U}$ and $p\in\mathcal{P}\textsubscript{warm}$; and (ii) a mapping function $\mathcal{M}(f\textsubscript{p}):\mathbb{R}\textsuperscript{\it I}\rightarrow\mathbb{R}\textsuperscript{\it D}$. This function is used for finding embeddings of products that have no purchase history. The learned embeddings are expected to represent characteristic features of users and products which can be exploited for making representations. The VASG framework achieves the above mentioned objectives in two steps- 
\begin{itemize}
    \item The first step involves training a Skip-Gram model on sequences of users and products. Here, the Skip-Gram architecture is modified by adding an auxiliary objective of learning an inverse mapping function $\mathcal{M}\textsuperscript{-1}(\mathcal{X}\textsubscript{p}): \mathbb{R}\textsuperscript{\it D}\rightarrow\mathbb{R}\textsuperscript{\it I}$. The {\it D} dimensional embeddings corresponding to users $u\in\mathcal{U}$ and products $p\in\mathcal{P}\textsubscript{warm}$ are learned during this step.
    \item The second step involves learning the mapping function $\mathcal{M}(f\textsubscript{p}):\mathbb{R}\textsuperscript{\it I}\rightarrow\mathbb{R}\textsuperscript{\it D}$. This is achieved by training a deep encoder-decoder architecture using the inverse mapping $\mathcal{M}\textsuperscript{-1}(.)$ learned in the previous step.
\end{itemize}
We describe the network structure used for representing user-product interactions before describing the model architecture. 

\subsection{Heterogeneous Information Network}

Skip-Gram \cite{DBLP:journals/corr/abs-1301-3781} is an unsupervised representation learning model for words in a text corpus. The representation corresponding to each word is learned by maximizing the probability of correctly predicting the context surrounding the given word. The proposed framework builds upon the Skip-Gram architecture, hence, it requires sequences of user and products. Taking inspiration from \cite{8355676}, we generate these sequences by performing random walks on a bipartite Heterogeneous Information Network (HIN) of users and products.

Formally, our HIN is a graph $G = (V,E)$ where $V = \mathcal{U}\cup\mathcal{P}\textsubscript{warm}$ is the set of nodes. $E = \{(u,p) \vert p\in\mathcal{P}\textsubscript{u}\cap\mathcal{P}\textsubscript{warm}~;~u\in\mathcal{U}\}$ is the set of edges connecting the nodes where each edge connects a user to the products purchased by him. Sequences are generated in the following manner - 
Starting from a node $\mathrm{v}\in V$, $n$ number of random walks, each of length $wl$ are performed on the HIN. Every subsequent node in a walk is chosen randomly from the direct neighbours of the current node. For example, consider the sample HIN shown in Figure 1(a). A sequence starting from node {\it u\textsubscript{1}} could be {\it u\textsubscript{1} p\textsubscript{2} u\textsubscript{2} p\textsubscript{2} u\textsubscript{3} ...} repeated for $wl$ nodes. Here, $n$ and $wl$  are hyper-parameters which determine the size of the corpus generated for training. It is to be noted that the bipartite structure of the HIN ensures that users and products appear in an alternating fashion in each random walk. The neighbouring context of a node, $\mathcal{N}(\mathrm{v})$ is determined by sliding a fixed size context window over the generated sequences. We discuss the effect of the context window size on the learned representations in Section 6.

\begin{table}[]
\caption{Dataset statistics (after preprocessing)}
\begin{tabular}{@{}llll@{}}
\toprule
\textbf{Dataset} & \textbf{\#products} & \textbf{\#users} & \textbf{\#ratings} \\ \midrule
Women            & 116,971             & 14,110           & 307,863            \\
Men              & 62,376              & 19,618           & 187,181            \\
Shoes            & 71,219              & 22,066           & 211,818            \\
Jewelry          & 62,958              & 13,975           & 161,439            \\ \bottomrule
\end{tabular}
\end{table}

\subsection{Visually Aware Skip-Gram}
Consider an HIN node $\mathrm{v}\in V$ and its neighbouring context  $\mathcal{N}(\mathrm{v})$. We can learn the the representations corresponding to each node in the HIN by optimizing the Skip-Gram objective - 
\begin{equation}
   \underset{\theta}{arg\,max} \displaystyle\sum_{\mathrm{v}\in V} \displaystyle\sum_{c \in N(\mathrm{v})} log (p(c|v;\theta))
\end{equation}
Here $p(c|v;\theta)$ is the commonly used softmax probability, given by: $p(c|v, \theta) = \frac{e^{\mathcal{X}\textsubscript{c}\mathcal{X}\textsubscript{v}}}{\sum_{n \in V} e^{\mathcal{X}\textsubscript{\it n}\mathcal{X}\textsubscript{v}}}$ where $\mathcal{X}\textsubscript{v}\in\mathbb{R}\textsuperscript{D}$ is the $D$ dimensional latent representation corresponding to node $\mathrm{v}$. Mikolov et al. \cite{Mikolov:2013:DRW:2999792.2999959} introduces negative sampling for efficiently estimating the softmax probabilities however, generating effective negative samples can be an expensive process \cite{10.1145/3289600.3290979}. We expect the negative samples to highlight the contrast in a product's visual appearance and a user's purchase preference simultaneously. In our experiments, we are able to obtain meaningful results without negative sampling, hence we approximate the softmax probability as:  $\sigma(\mathcal{X}\textsubscript{c}~.~\mathcal{X}\textsubscript{v})$ where $\sigma$ is the $sigmoid$ function, given by: $\sigma(x) = \frac{1}{1+e^{-x}}$.

We add the auxiliary objective of learning $\mathcal{M}\textsuperscript{-1}(\mathcal{X}\textsubscript{p}): \mathbb{R}\textsuperscript{\it D}\rightarrow\mathbb{R}\textsuperscript{\it I}$ to the Skip-Gram objective whenever $\mathrm{v}\in\mathcal{P}\textsubscript{warm}$. The function $\mathcal{M}\textsuperscript{-1}(.)$ attempts to reconstruct the DeepCNN image feature $f\textsubscript{p}$ of a product from its latent representation $\mathcal{X}\textsubscript{p}$. This inverse mapping function is approximated by training a decoder architecture with the objective of minimizing the mean squared error of image feature reconstruction - 

\begin{equation}
   \underset{\mathcal{M}^{-1}}{arg\,min} \displaystyle\sum_{\mathrm{v}\in \mathcal{P}\textsubscript{warm}}  
   \vert\vert f\textsubscript{v} - f^{'}_{\textsubscript{p}}\vert\vert^{2}_{2}
\end{equation}

Where $f^{'}_{\textsubscript{p}} = \mathcal{M}\textsuperscript{-1}(\mathcal{X}\textsubscript{p})$. It is to be noted that the auxiliary task is optimized only when $\mathrm{v}\in\mathcal{P}\textsubscript{warm}$. This implies that users have a single loss function while products have two different loss functions which are optimized simultaneously in a multi-task learning fashion. When a node $\mathrm{v}$ is observed in the training corpus, the loss can be written as - 

\begin{equation}
    \mathcal{L}(\mathrm{v})=
    \begin{cases}
    
    -log(\sigma(\mathcal{X}\textsubscript{c}\mathcal{X}\textsubscript{v}))
    &\text{ $\mathrm{v}\in \mathcal{U}$}\\
    
    w\textsubscript{1}(-log(\sigma(\mathcal{X}\textsubscript{c}\mathcal{X}\textsubscript{v}))) +
    w\textsubscript{2}\vert\vert f\textsubscript{v} - f^{'}_{\textsubscript{p}}\vert\vert^{2}_{2}
    & \text{$\mathrm{v}\in \mathcal{P}\textsubscript{warm}$}
    
    \end{cases}
\end{equation}

Here  $ w\textsubscript{1} $ and $ w\textsubscript{2} $ are trainable weights used for combining the Skip-Gram and image reconstruction losses. We follow the approach introduced in \cite{DBLP:journals/corr/KendallGC17} for learning these weights.

Once the embeddings and the inverse mapping $\mathcal{M}^{-1}(.)$ have been learned, the function $\mathcal{M}(.)$ is approximated by training a deep autoencoder for reconstructing the product image features for each product $p\in\mathcal{P}\textsubscript{warm}$. The autoencoder has $\mathcal{M}^{-1}(.)$ as the decoder while the encoder attempts to approximate $\mathcal{M}(.)$. Weight updates are not allowed in the decoder, therefore $\mathcal{M}^{-1}(.)$ remains unchanged during the training process. An overview of the proposed architecture is presented in Figure 1. 
\\{\bfseries Implementation Details:}
We implemented our model in Pytorch. The decoder architecture $\mathcal{M}^{-1}(.)$ consists of 5 fully connected layers with [256, 512, 1024, 2048, $I$] neurons respectively ($I = 4096$). Each layer has {\it reLu} activation and dropout regularization with dropout probability set to 0.5. The encoder $\mathcal{M}(.)$ follows the same architecture. Adam optimizer with learning rate 1e-3 is used for parameter updates.
To fully utilize GPU acceleration, we use a batch wise implementation and ensure each batch homogeneously comprises either users or products as we iterate over the Skip-Gram corpus. Training on our largest dataset requires around 30 minutes on a Nvidia Quadro P5000 GPU. We provide our implementation here\footnote{\href{https://github.com/parth2170/triplet-recsys}{\bfseries Link to repository}. Will be released at the time of publication.}

\setlength{\textfloatsep}{0.3cm}

\section{Experiments}
We perform experiments on real-world datasets to evaluate the VASG embeddings. Our experiments attempt to answer the following research questions - 
{\bf (i)} Can the embeddings generate personalized recommendations for each user?
{\bf (ii)} Can the embeddings identify product to product relationships?
{\bf (iii)} What are the properties displayed by the learned embedding space?
{\bf (vi)} How do different components and hyper-parameters of the proposed framework affect its performance?

\subsection{Datasets}
Four subcategories under the "Clothing, Shoes \& Jewelry" dataset from {\it Amazon Product Reviews} \cite{McATarShiHen15} are used for our experiments. The selected subcategories are - {\it Men, Women, Shoes} and {\it Jewelry}. The dataset is pre-processed to remove users that do not have sufficient purchase history. Users with $\vert\mathcal{P}\textsubscript{u}\vert<5$ are discarded from the datasets . Users' rating history and product image features are used in our framework. Product metadata is also available, however it not utilized for learning embeddings. Only image URLs are used for retrieving images for the purpose of visualization. Product image features were extracted using a pre-trained Caffe reference model \cite{10.1145/2647868.2654889}, with 5 convolutional layers followed by 3 dense layers. The feature vector is obtained by taking the output of the second dense layer (FC7) of this network and has length $I = 4096$. The details   of the datasets are specified in Table 1.

The training and test sets are created by following the leave one out protocol described by \cite{10.5555/3015812.3015834}. For each user $u$, one random product rated by the user is used for testing, while the remaining products are used for training. Two subsets of the test set are created for testing in {\it warm start} and {\it completely cold start} setting:
$\mathcal{T}\textsubscript{warm} = \{(u, p) \mid u \in \mathcal{U}~;~p\in \mathcal{P}\textsubscript{warm}\}$ and $\mathcal{T}\textsubscript{cold} = \{(u, p) \mid u \in \mathcal{U}~;~ p\in \mathcal{P}\textsubscript{cold}\}$. Here $\mathcal{T}\textsubscript{warm}$ and $\mathcal{T}\textsubscript{cold}$ are approximately equal in size.

\begin{table}[t]
\caption{AUC values on the test sets. The best scores are boldfaced. Refer Section 4.2}
\begin{tabular}{@{}llllll@{}}
\toprule
\textbf{Dataset} & \textbf{Setting}       & \textbf{RAND} & \textbf{WBOI} & \textbf{VBPR} & \textbf{VASG}  \\ \midrule
Women            & \textit{Warm Products}  & 0.4998        & 0.6134        & 0.7982        & \textbf{0.8964} \\
                 & \textit{Cold Products} & 0.5001        & 0.5912        &               & \textbf{0.7596} \\
Men              & \textit{Warm Products}  & 0.4897        & 0.6468        & 0.7753        & \textbf{0.8758} \\
                 & \textit{Cold Products} & 0.4954        & 0.6094        &               & \textbf{0.7061} \\
Shoes            & \textit{Warm Products}  & 0.5023        & 0.6318        & 0.8116        & \textbf{0.8972} \\
                 & \textit{Cold Products} & 0.4995        & 0.5846        &               & \textbf{0.7261} \\
Jewelry          & \textit{Warm Products}  & 0.4987        & 0.6087        & 0.7629        & \textbf{0.8801} \\
                 & \textit{Cold Products} & 0.5003        & 0.5704        &               & \textbf{0.7134} \\ \bottomrule
\end{tabular}
\end{table}

\subsection{Making Personalized Recommendations}

VASG embeddings can be used for generating recommendations by searching for relevant products in the vicinity of corresponding users in the learned latent space. We evaluate this ability to generate personalized recommendations in a Bayesian Personalized Ranking (BPR) scenario. BPR based methods \cite{10.5555/3015812.3015834,DBLP:journals/corr/abs-1205-2618} learn a rating prediction function by optimizing pairwise rankings of products w.r.t. to users.
Predicted rankings are then evaluated using the well known AUC (Area Under ROC Curve) metric:

\begin{equation}
  AUC = \frac{1}{\vert\mathcal{T}\textsubscript{test}\vert}\sum_{{}u\in\mathcal{T}\textsubscript{test}}\frac{1}{\vert\mathcal{P}-\mathcal{P}\textsubscript{u}\vert} \sum_{{}p\textsubscript{j}\in \mathcal{P}-\mathcal{P}\textsubscript{u}}\textbf{1}((u, p\textsubscript{test}) > (u,p\textsubscript{j}))
\end{equation}

where $\mathcal{T}\textsubscript{test}$ corresponds to $\mathcal{T}\textsubscript{warm}$ or $\mathcal{T}\textsubscript{cold}$ and $u,p\textsubscript{test}\in\mathcal{T}\textsubscript{test}$. Here, $\textbf{1}(.)$ is an indicator function which evaluates if $p\textsubscript{test}$ has been ranked higher than $p\textsubscript{j}$ for user $u$. BPR based methods rank products using the learned rating prediction function. We evaluate VASG embeddings for ranking products using the {\it cosine similarity} between corresponding user and product embeddings.

\subsubsection{{\bfseries Baselines}}

We  compare the proposed embeddings against the raw image features and against a state-of-the-art visually aware BPR method:
\begin{itemize}
    \item {\bfseries RAND} ({\it Random}) - Product rankings are decided randomly for all users.
    \item {\bfseries WBOI} ({\it Weighted Bag Of Images}) - User embeddings are directly computed using the image features corresponding to the products purchased by him. The rating weighted mean of image features is used as the user embedding and ranks are computed using cosine similarity as described above.
    \item {\bfseries VBPR} ({\it Visual Bayesian Product Ranking}) - Introduced by \cite{10.5555/3015812.3015834}, VBPR incorporates visual factors to the BPR framework by using product image DeepCNN features in its rating prediction function.
\end{itemize}

\subsubsection{{\bfseries Results}}
The entire pipeline for VASG with $D = 100$ is run 5 times and the averaged results are reported in Table 2. The results for VBPR are calculated using 20 latent factors and 100 visual factors.
Results are reported on the set $\mathcal{T}\textsubscript{warm}$ ({\it Warm Products}) and for on $\mathcal{T}\textsubscript{cold}$ ({\it Cold Products}) separately. VASG embeddings for cold start products are found using the mapping $\mathcal{M}(.)$.
Since latent factors for cold start products are not available in VBPR, computing its results on the set $\mathcal{T}\textsubscript{cold}$ is infeasible. 

VASG embeddings show an average improvement of 3.84\% over VBPR in the warm start setting. The AUC scores drop in the cold start setting, however they are significantly better than the WBOI baseline. This drop in performance can be explained by imperfections in the approximation of the mapping function $\mathcal{M}(.)$.

\subsection{Identifying Product Relationships}
We expect the embeddings corresponding to co-purchased products to be similar. The evaluation strategy used by McAuley et al. (IBR) \cite{McATarShiHen15} is followed for identifying product relationships using VASG embeddings. IBR uses product DeepCNN image features to learn a parametric distance function by maximizing the probability of correctly identifying co-purchased product pairs. Formally, two products $p\textsubscript{i}$ and $p\textsubscript{j}$ share a relation $R\textsubscript{ij}$ if they have been purchased by the same user. This relation is termed as an {\it "also bought"} relation in IBR. We attempt to correctly identify co-purchased product pairs using the cosine similarity between the VASG embeddings. The test set is restructured in the following manner for reporting the results - A set of positive relation pairs $\mathcal{R} = \{(p\textsubscript{i}, p\textsubscript{j}) \mid \exists~R\textsubscript{ij}\}$ and a set of negative relation pairs $\mathcal{Q} = \{(p\textsubscript{i}, p\textsubscript{j}) \mid \nexists~R\textsubscript{ij}\}$ is constructed such that $\vert\mathcal{R}\vert = \vert\mathcal{Q}\vert~;~p\textsubscript{i}\in\mathcal{P}$. A positive relation is identified when 
$cosine~similarity(\mathcal{X}\textsubscript{i}, \mathcal{X}\textsubscript{j}) > t$, where $\mathcal{X}\textsubscript{i}$ is the VASG embedding corresponding to $p\textsubscript{i}$ and $t$ is a manually assigned threshold. The accuracy of identifying co-purchased product pairs is reported on the combined set $\mathcal{R}\cup \mathcal{Q}$.

\begin{table}[t]
\caption{Accuracy of predicting product relations. Refer Section 4.3}
\begin{tabular}{@{}llllll@{}}
\toprule
\textbf{Dataset} & \textbf{INN} & \textbf{IBR} & \textbf{VASG} & \begin{tabular}[c]{@{}l@{}}VASG \\ vs INN\end{tabular} & \begin{tabular}[c]{@{}l@{}}VASG \\ vs IBR\end{tabular} \\ \midrule
Women            & 67.82\%      & 87.43\%      & 79.59\%        & 17\%                                                    & -10\%                                                   \\
Men              & 66.23\%      & 87.13\%      & 80.12\%        & 21\%                                                    & -9\%                                                    \\
Shoes            & 69.34\%      & 89.67\%      & 81.56\%        & 18\%                                                    & -10\%                                                   \\
Jewelry          & 65.42\%      & 83.25\%      & 79.36\%        & 21\%                                                    & -5\%                                                    \\ \bottomrule
\end{tabular}
\end{table}

\begin{figure*}[]
  \centering
  \includegraphics[width=\linewidth]{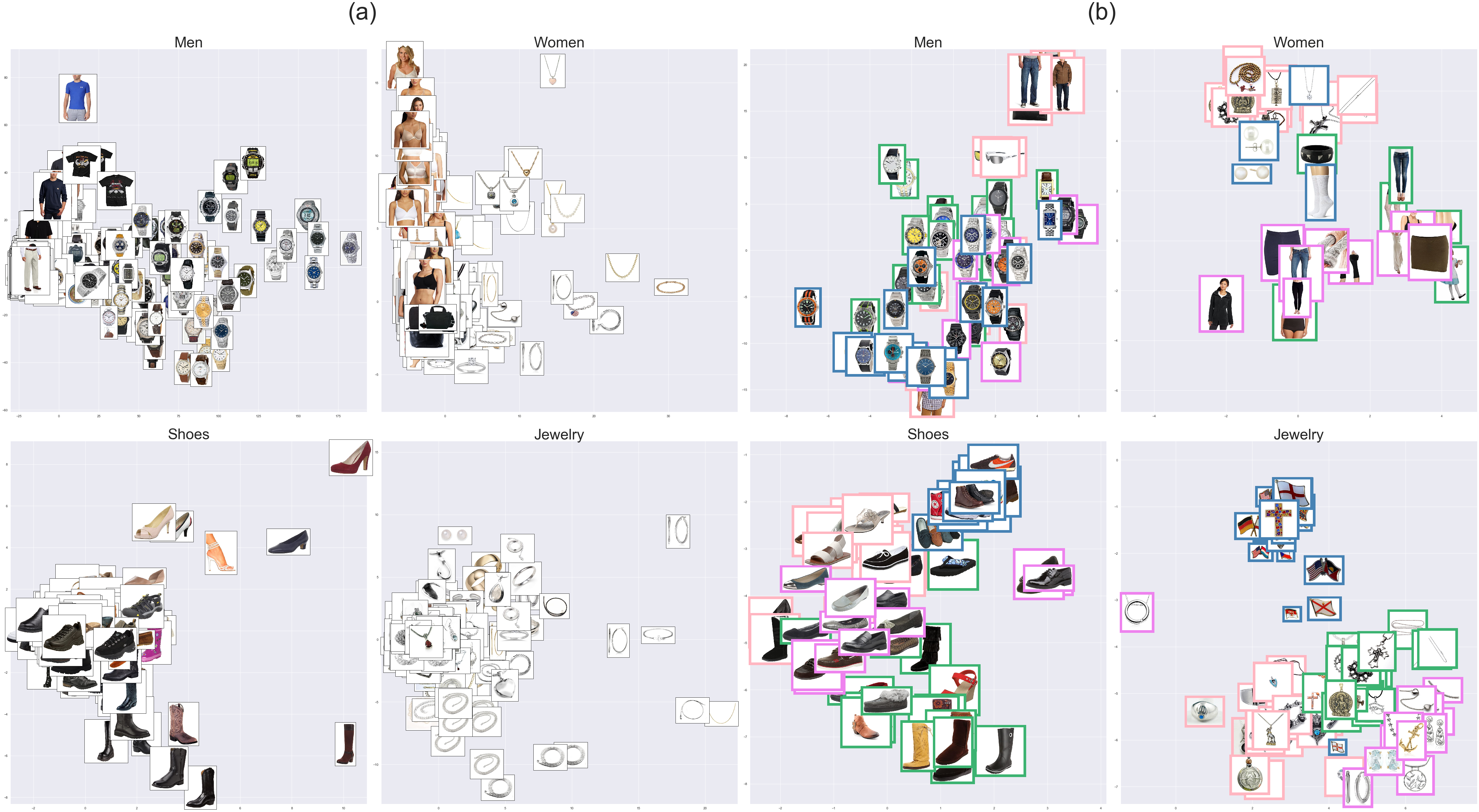}
  \caption{2D visualizations of VASG embeddings(a) PCA projections of 100 dimensional embeddings for each dataset. Here 200 randomly selected products are plotted (b) t-SNE projections of 100 dimensional embeddings for each dataset. Here products purchased by 4 randomly selected users are plotted. Products corresponding to the same user have the same frame colour}
\end{figure*}

\subsubsection{{\bfseries Baselines}}

Since $\vert\mathcal{R}\vert = \vert\mathcal{Q}\vert$, random classification is 50\% accurate. In addition, the performance of VASG embeddings is compared to the following methods:
\begin{itemize}
    \item {\bfseries INN} ({\it Image Nearest Neighbours}) - Cosine similarity between the DeepCNN image features $f\textsubscript{i}, f\textsubscript{j}$ are used for identifying relationships in the same manner as described above.
    \item {\bfseries IBR} ({\it Image Based Recommendations}) - As described above, this method is trained on all co-purchased product pairs occurring in the training set.
\end{itemize}

\subsubsection{{\bfseries Results}}
The entire pipeline for VASG with $D = 100$ is run 5 times and the averaged results are reported in Table 3. For IBR, the rank of the Mahalanobis transform is set to 100 for fair comparison. Here, VASG embeddings show an average improvement of 19\% over INN, however they do not outperform IBR which shows significantly better results across all datasets. 
However it should be noted that IBR is trained for specifically identifying all product-product relations \cite{McATarShiHen15} whereas VASG embeddings are not. Our method learns embeddings by capturing co-occurrence of users and products in a small network neighbourhood which is more suited for making personalised recommendations which is shown by the results in Section 4.2 

\section{Analysis of Embeddings}

{\bfseries Embedding Visualization.}
We visualize the learned embeddings in lower dimensions to understand the information captured. We plot 2D projections of a random set of products from each dataset in Figure 2(a). Here dimensions are reduced using Principal Component Analysis (PCA). A variance in the visual appearance of the products can be seen when we travel along the axes in these plots. For example in the {\it Men} dataset, the x-axis is populated with watches while the y-axis contains clothing related items like t-shirts. A similar trend can be observed for the {\it Women} and {\it Shoes} dataset, however, the plot for the {\it Jewelry} dataset looks noisy.
Next, we visualize groups of products which have been purchased by specific users in Figure 2(b). Here we select 4 random users and plot all the products purchased by them. Dimensions are reduced using t-SNE with the perplexity parameter set to 30. Products corresponding to a specific user are marked by the colour of the frame around them. In case of co-purchased products, the frame colour is chosen randomly. It can be seen that products purchased by the same user have been clustered together. These plots show that the leaned embeddings can simultaneously capture the visual properties and the purchase history of a product. \\
\begin{figure}[t]
  \includegraphics[width=\linewidth]{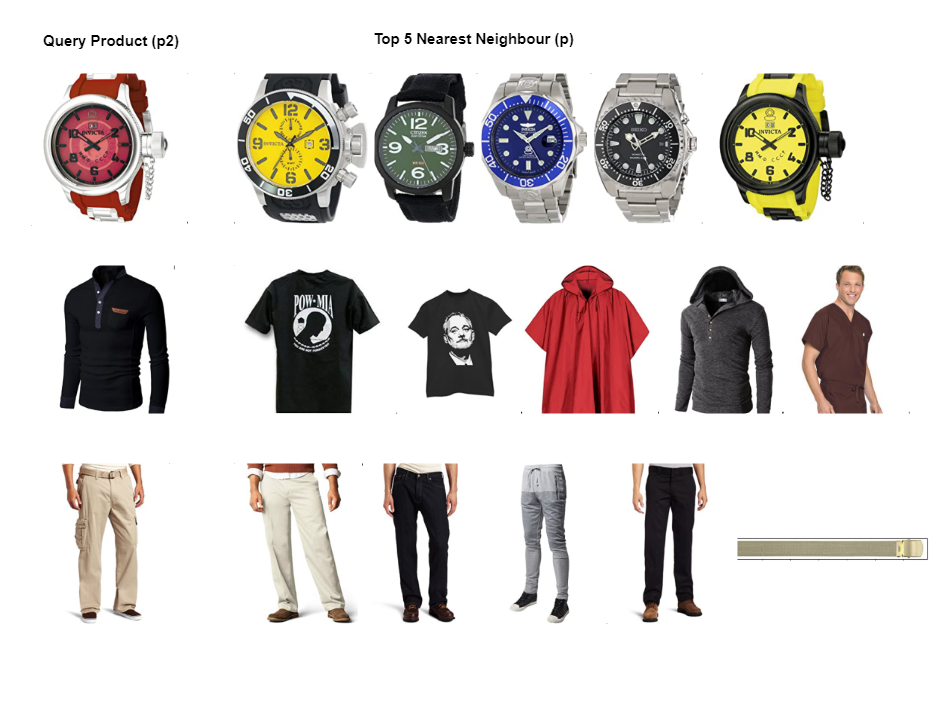}
  \caption{Examples of recommendations generated by exploiting linear relationships in the embeddings}
  \Description{A woman and a girl in white dresses sit in an open car.}
\end{figure}
{\bfseries Linear relationships.}
Skip-Gram word embeddings are known to follow a linear algebraic structure where embeddings display associative properties (shown by the famous example vector(”King”) - vector(”Man”) + vector(”Woman”)$\approx$vector(”Queen”) \cite{DBLP:journals/corr/abs-1301-3781}). We attempt to exploit a similar structure in the VASG embeddings for recommending products. Consider
a pair:  $\{(u1,p1), p1 \in \mathcal{P}\textsubscript{u1}\}$ and a user $u2;u2\neq u1$, keeping $(u1,p1)$ and $u2$ fixed,
if we can a find a product $ p \in \mathcal{P}$ such that $$ cosine~similarity(\mathcal{X}\textsubscript{p1}-\mathcal{X}\textsubscript{u1}+\mathcal{X}\textsubscript{u2},\mathcal{X}\textsubscript{p}) \approx 1 $$ Then we say that $p$ is a user specific recommendation generated for the user $u2$ having a query product $p1$. To test this methodology, we sample 10,000 pairs of $u1,u2$ and find the nearest neighbour of the point $\mathcal{X}\textsubscript{p1}+\mathcal{X}\textsubscript{u2}-\mathcal{X}\textsubscript{u1}$, using cosine  distance, from the set $\mathcal{P}$ and treat them as recommendations generated by VASG. We obtained a  precision@5 of 89.7\% in the Men dataset. Figure 3. shows  some of the recommendations generated by this methodology. 
Approximately speaking, here a product $p$ is recommended for user $u2$ when the similarity between $p,p1$ and between $p,u2$ is high and when similarity between $u1,p$ is low.\\
{\bfseries Embedding similarity distribution.}
This analysis is inspired by \cite{10.1145/3219819.3219890}. Effectiveness of the learned embeddings can be judged by the distribution of distances between random pairs of embeddings. A wide distribution indicates that the embedding space has enough “resolution” to capture relevance of different product pairs. Figure 4 plots the distribution of cosine similarities between pairs of products from the {\it Men } dataset. The distribution coming from the VASG embeddings is compared against the distribution computed from product image features. VASG embeddings show a wider distribution hence proving their superiority over using raw image features. The kurtosis of this distribution for VASG embeddings is 0.87, compared to 3.67 for image features.

\begin{figure}[t]
  \includegraphics[width=\linewidth]{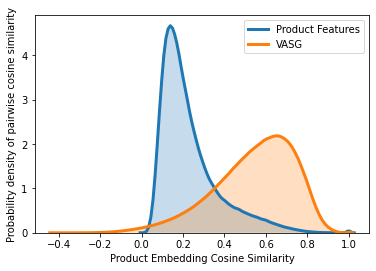}
  \caption{Probability density of pairwise cosine similarity obtained from VASG embeddings and Product (image) Features.}
  \Description{A woman and a girl in white dresses sit in an open car.}
\end{figure}
\section{Sensitivity Analysis}

{\bfseries Influence of Embedding size.} We study the variation of AUC scores with the embedding dimension size $D$. Figure 4 shows AUC scores for warm-start products when evaluating the performance of making personalized recommendations. The performance improves with increase in dimension size, however the rate of improvement seems to slow down after $D = 70$.\\
{\bfseries Influence of noise addition.} Autoencoders are known to suffer from over fitting. To overcome this problem, we add a small amount of Gaussian noise to the product image features while learning the mapping $\mathcal{M}(.)$. We find that noise addition helps in improving the performance of cold-start products for making personalized recommendations by an average margin of 1.8\% in AUC score.\\
{\bfseries Influence of Skip-Gram window Size.} 
The window size determines the neighbourhood size of a HIN node. For example a window size of 5 restricts the neighborhood to second degree neighbours. A smaller context window exposes the embeddings to local HIN structures which is suited for making personalized recommendations. We experiment with window size [3,5,7,9] and find that a window size of 7 gives the best results. Smaller window sizes lead to a faster convergence, but the embeddings over-fit in a small neighbourhood whereas larger window sizes make the training corpus unnecessarily large.\\
{\bfseries Influence of the auxiliary decoder.} 
We study how VASG's performance is affected by removing the auxiliary decoder. This reduces the proposed architecture to the general Skip-Gram model. We evaluate the performance of making personalized recommendations in the set $\mathcal{T}\textsubscript{warm}$. The performance drops by an average of 2.4\% on the {\it Women, Men} and {\it Shoes} dataset while the performance remains almost consistent for the {\it Jewelry} dataset. This indicates that addition of visual features improves the convential Skip-Gram embeddings.

\begin{figure}[t]
  \centering
  \includegraphics[width=\linewidth]{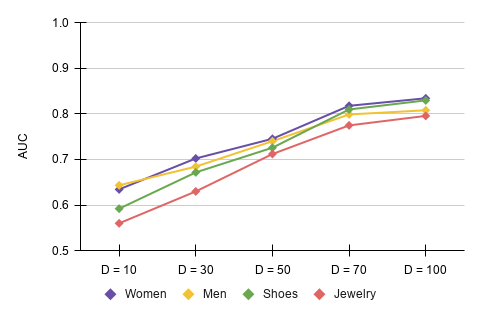}
  \caption{Variation of AUC scores with embedding dimensions}
\end{figure}

\section{Conclusion}
In this paper, we proposed a novel embedding learning architecture, VASG, which incorporates visual features into the Skip-Gram model. This model has two different loss functions depending on whether a user or a product is observed during training. VASG embeddings can be used for making personalized recommendations using nearest neighbour searches in the learned latent space. In addition, this framework enables us to find embeddings for cold start products which have never been observed during training. The embeddings are trained on real world datasets and extensive analysis brings forward several interesting properties captured by them.

There are multiple potential directions in which this work can be extended. Use of effective negative samples while training can enable us to learn a more refined latent space. Adversarial training can be used to reconstruct product images using Generative Adversarial Networks conditioned on the Skip-Gram embeddings. Additional auxiliary tasks which bring information from other modalities can be added to central Skip-Gram architecture. VASG embeddings can be also be incorporated into frameworks like SVDFeature for predicting ratings. 


\bibliographystyle{ACM-Reference-Format}
\bibliography{paper}


\begin{thebibliography}{37}


\ifx \showCODEN    \undefined \def \showCODEN     #1{\unskip}     \fi
\ifx \showDOI      \undefined \def \showDOI       #1{#1}\fi
\ifx \showISBNx    \undefined \def \showISBNx     #1{\unskip}     \fi
\ifx \showISBNxiii \undefined \def \showISBNxiii  #1{\unskip}     \fi
\ifx \showISSN     \undefined \def \showISSN      #1{\unskip}     \fi
\ifx \showLCCN     \undefined \def \showLCCN      #1{\unskip}     \fi
\ifx \shownote     \undefined \def \shownote      #1{#1}          \fi
\ifx \showarticletitle \undefined \def \showarticletitle #1{#1}   \fi
\ifx \showURL      \undefined \def \showURL       {\relax}        \fi
\providecommand\bibfield[2]{#2}
\providecommand\bibinfo[2]{#2}
\providecommand\natexlab[1]{#1}
\providecommand\showeprint[2][]{arXiv:#2}

\bibitem[\protect\citeauthoryear{Chen, Zhang, He, Nie, Liu, and Chua}{Chen
  et~al\mbox{.}}{2017}]%
        {10.1145/3077136.3080797}
\bibfield{author}{\bibinfo{person}{Jingyuan Chen}, \bibinfo{person}{Hanwang
  Zhang}, \bibinfo{person}{Xiangnan He}, \bibinfo{person}{Liqiang Nie},
  \bibinfo{person}{Wei Liu}, {and} \bibinfo{person}{Tat-Seng Chua}.}
  \bibinfo{year}{2017}\natexlab{}.
\newblock \showarticletitle{Attentive Collaborative Filtering: Multimedia
  Recommendation with Item- and Component-Level Attention}. In
  \bibinfo{booktitle}{\emph{Proceedings of the 40th International ACM SIGIR
  Conference on Research and Development in Information Retrieval}} (Shinjuku,
  Tokyo, Japan) \emph{(\bibinfo{series}{SIGIR ’17})}.
  \bibinfo{publisher}{Association for Computing Machinery},
  \bibinfo{pages}{335–344}.
\newblock
\showISBNx{9781450350228}


\bibitem[\protect\citeauthoryear{Chen, Zhang, Lu, Chen, Zheng, and Yu}{Chen
  et~al\mbox{.}}{2012}]%
        {Chen:2012:STF:2503308.2503357}
\bibfield{author}{\bibinfo{person}{Tianqi Chen}, \bibinfo{person}{Weinan
  Zhang}, \bibinfo{person}{Qiuxia Lu}, \bibinfo{person}{Kailong Chen},
  \bibinfo{person}{Zhao Zheng}, {and} \bibinfo{person}{Yong Yu}.}
  \bibinfo{year}{2012}\natexlab{}.
\newblock \showarticletitle{SVDFeature: A Toolkit for Feature-based
  Collaborative Filtering}.
\newblock \bibinfo{journal}{\emph{J. Mach. Learn. Res.}} \bibinfo{volume}{13},
  \bibinfo{number}{1} (\bibinfo{date}{Dec.} \bibinfo{year}{2012}),
  \bibinfo{pages}{3619--3622}.
\newblock
\showISSN{1532-4435}


\bibitem[\protect\citeauthoryear{Chen, Chen, Xu, Zhang, Cao, Qin, and Zha}{Chen
  et~al\mbox{.}}{2019}]%
        {10.1145/3331184.3331254}
\bibfield{author}{\bibinfo{person}{Xu Chen}, \bibinfo{person}{Hanxiong Chen},
  \bibinfo{person}{Hongteng Xu}, \bibinfo{person}{Yongfeng Zhang},
  \bibinfo{person}{Yixin Cao}, \bibinfo{person}{Zheng Qin}, {and}
  \bibinfo{person}{Hongyuan Zha}.} \bibinfo{year}{2019}\natexlab{}.
\newblock \showarticletitle{Personalized Fashion Recommendation with Visual
  Explanations Based on Multimodal Attention Network: Towards Visually
  Explainable Recommendation}. In \bibinfo{booktitle}{\emph{Proceedings of the
  42nd International ACM SIGIR Conference on Research and Development in
  Information Retrieval}} (Paris, France)
  \emph{(\bibinfo{series}{SIGIR’19})}. \bibinfo{publisher}{Association for
  Computing Machinery}, \bibinfo{pages}{765–774}.
\newblock
\showISBNx{9781450361729}


\bibitem[\protect\citeauthoryear{Cheng, Chang, Zhu, Kanjirathinkal, and
  Kankanhalli}{Cheng et~al\mbox{.}}{2019}]%
        {10.1145/3291060}
\bibfield{author}{\bibinfo{person}{Zhiyong Cheng}, \bibinfo{person}{Xiaojun
  Chang}, \bibinfo{person}{Lei Zhu}, \bibinfo{person}{Rose~C. Kanjirathinkal},
  {and} \bibinfo{person}{Mohan Kankanhalli}.} \bibinfo{year}{2019}\natexlab{}.
\newblock \showarticletitle{MMALFM: Explainable Recommendation by Leveraging
  Reviews and Images}.
\newblock \bibinfo{journal}{\emph{ACM Trans. Inf. Syst.}} \bibinfo{volume}{37},
  \bibinfo{number}{2}, Article \bibinfo{articleno}{16} (\bibinfo{date}{Jan.}
  \bibinfo{year}{2019}).
\newblock
\showISSN{1046-8188}


\bibitem[\protect\citeauthoryear{Dong, Chawla, and Swami}{Dong
  et~al\mbox{.}}{2017}]%
        {dong2017metapath2vec}
\bibfield{author}{\bibinfo{person}{Yuxiao Dong}, \bibinfo{person}{Nitesh~V
  Chawla}, {and} \bibinfo{person}{Ananthram Swami}.}
  \bibinfo{year}{2017}\natexlab{}.
\newblock \showarticletitle{metapath2vec: Scalable Representation Learning for
  Heterogeneous Networks}. In \bibinfo{booktitle}{\emph{KDD '17}}. ACM,
  \bibinfo{pages}{135--144}.
\newblock


\bibitem[\protect\citeauthoryear{{Gajic} and {Baldrich}}{{Gajic} and
  {Baldrich}}{2018}]%
        {8575408}
\bibfield{author}{\bibinfo{person}{B. {Gajic}} {and} \bibinfo{person}{R.
  {Baldrich}}.} \bibinfo{year}{2018}\natexlab{}.
\newblock \showarticletitle{Cross-Domain Fashion Image Retrieval}. In
  \bibinfo{booktitle}{\emph{2018 IEEE/CVF Conference on Computer Vision and
  Pattern Recognition Workshops (CVPRW)}}. \bibinfo{pages}{1950--19502}.
\newblock


\bibitem[\protect\citeauthoryear{{Gantner}, {Drumond}, {Freudenthaler},
  {Rendle}, and {Schmidt-Thieme}}{{Gantner} et~al\mbox{.}}{2010}]%
        {5693971}
\bibfield{author}{\bibinfo{person}{Z. {Gantner}}, \bibinfo{person}{L.
  {Drumond}}, \bibinfo{person}{C. {Freudenthaler}}, \bibinfo{person}{S.
  {Rendle}}, {and} \bibinfo{person}{L. {Schmidt-Thieme}}.}
  \bibinfo{year}{2010}\natexlab{}.
\newblock \showarticletitle{Learning Attribute-to-Feature Mappings for
  Cold-Start Recommendations}. In \bibinfo{booktitle}{\emph{2010 IEEE
  International Conference on Data Mining}}. \bibinfo{pages}{176--185}.
\newblock


\bibitem[\protect\citeauthoryear{Gidaris, Bursuc, Komodakis, Perez, and
  Cord}{Gidaris et~al\mbox{.}}{2019}]%
        {Gidaris_2019_ICCV}
\bibfield{author}{\bibinfo{person}{Spyros Gidaris}, \bibinfo{person}{Andrei
  Bursuc}, \bibinfo{person}{Nikos Komodakis}, \bibinfo{person}{Patrick Perez},
  {and} \bibinfo{person}{Matthieu Cord}.} \bibinfo{year}{2019}\natexlab{}.
\newblock \showarticletitle{Boosting Few-Shot Visual Learning With
  Self-Supervision}. In \bibinfo{booktitle}{\emph{Proceedings of the IEEE/CVF
  International Conference on Computer Vision (ICCV)}}.
\newblock


\bibitem[\protect\citeauthoryear{Gong, Lin, Song, and Wang}{Gong
  et~al\mbox{.}}{2020}]%
        {10.1145/3336191.3371770}
\bibfield{author}{\bibinfo{person}{Lin Gong}, \bibinfo{person}{Lu Lin},
  \bibinfo{person}{Weihao Song}, {and} \bibinfo{person}{Hongning Wang}.}
  \bibinfo{year}{2020}\natexlab{}.
\newblock \showarticletitle{JNET: Learning User Representations via Joint
  Network Embedding and Topic Embedding}. In
  \bibinfo{booktitle}{\emph{Proceedings of the 13th International Conference on
  Web Search and Data Mining}} (Houston, TX, USA) \emph{(\bibinfo{series}{WSDM
  ’20})}. \bibinfo{publisher}{Association for Computing Machinery},
  \bibinfo{pages}{205–213}.
\newblock
\showISBNx{9781450368223}


\bibitem[\protect\citeauthoryear{Grover and Leskovec}{Grover and
  Leskovec}{2016}]%
        {grover2016node2vec}
\bibfield{author}{\bibinfo{person}{Aditya Grover} {and} \bibinfo{person}{Jure
  Leskovec}.} \bibinfo{year}{2016}\natexlab{}.
\newblock \showarticletitle{node2vec: Scalable feature learning for networks}.
  In \bibinfo{booktitle}{\emph{Proceedings of the 22nd ACM SIGKDD international
  conference on Knowledge discovery and data mining}}.
  \bibinfo{pages}{855--864}.
\newblock


\bibitem[\protect\citeauthoryear{Guo, Cheng, Nie, Xu, and Kankanhalli}{Guo
  et~al\mbox{.}}{2018}]%
        {10.1145/3240508.3240541}
\bibfield{author}{\bibinfo{person}{Yangyang Guo}, \bibinfo{person}{Zhiyong
  Cheng}, \bibinfo{person}{Liqiang Nie}, \bibinfo{person}{Xin-Shun Xu}, {and}
  \bibinfo{person}{Mohan Kankanhalli}.} \bibinfo{year}{2018}\natexlab{}.
\newblock \showarticletitle{Multi-Modal Preference Modeling for Product
  Search}. In \bibinfo{booktitle}{\emph{Proceedings of the 26th ACM
  International Conference on Multimedia}} (Seoul, Republic of Korea)
  \emph{(\bibinfo{series}{MM ’18})}. \bibinfo{publisher}{Association for
  Computing Machinery}, \bibinfo{pages}{1865–1873}.
\newblock
\showISBNx{9781450356657}


\bibitem[\protect\citeauthoryear{He, Fang, Wang, and McAuley}{He
  et~al\mbox{.}}{2016a}]%
        {10.1145/2959100.2959152}
\bibfield{author}{\bibinfo{person}{Ruining He}, \bibinfo{person}{Chen Fang},
  \bibinfo{person}{Zhaowen Wang}, {and} \bibinfo{person}{Julian McAuley}.}
  \bibinfo{year}{2016}\natexlab{a}.
\newblock \showarticletitle{Vista: A Visually, Socially, and Temporally-Aware
  Model for Artistic Recommendation}. In \bibinfo{booktitle}{\emph{Proceedings
  of the 10th ACM Conference on Recommender Systems}} (Boston, Massachusetts,
  USA) \emph{(\bibinfo{series}{RecSys ’16})}. \bibinfo{publisher}{Association
  for Computing Machinery}, \bibinfo{pages}{309–316}.
\newblock
\showISBNx{9781450340359}


\bibitem[\protect\citeauthoryear{He, Lin, Wang, and McAuley}{He
  et~al\mbox{.}}{2016b}]%
        {HeLinWanMcA16}
\bibfield{author}{\bibinfo{person}{Ruining He}, \bibinfo{person}{Chunbin Lin},
  \bibinfo{person}{Jianguo Wang}, {and} \bibinfo{person}{Julian McAuley}.}
  \bibinfo{year}{2016}\natexlab{b}.
\newblock \showarticletitle{Sparse hierarchical embeddings for visually-aware
  one-class collaborative filtering}. In
  \bibinfo{booktitle}{\emph{International Joint Conference on Artificial
  Intelligence}}.
\newblock


\bibitem[\protect\citeauthoryear{He and McAuley}{He and McAuley}{2016}]%
        {10.5555/3015812.3015834}
\bibfield{author}{\bibinfo{person}{Ruining He} {and} \bibinfo{person}{Julian
  McAuley}.} \bibinfo{year}{2016}\natexlab{}.
\newblock \showarticletitle{VBPR: Visual Bayesian Personalized Ranking from
  Implicit Feedback}. In \bibinfo{booktitle}{\emph{Proceedings of the Thirtieth
  AAAI Conference on Artificial Intelligence}} (Phoenix, Arizona)
  \emph{(\bibinfo{series}{AAAI’16})}. \bibinfo{publisher}{AAAI Press},
  \bibinfo{pages}{144–150}.
\newblock


\bibitem[\protect\citeauthoryear{Jain, Balasubramanian, Chunduri, and
  Varma}{Jain et~al\mbox{.}}{2019}]%
        {10.1145/3289600.3290979}
\bibfield{author}{\bibinfo{person}{Himanshu Jain}, \bibinfo{person}{Venkatesh
  Balasubramanian}, \bibinfo{person}{Bhanu Chunduri}, {and}
  \bibinfo{person}{Manik Varma}.} \bibinfo{year}{2019}\natexlab{}.
\newblock \showarticletitle{Slice: Scalable Linear Extreme Classifiers Trained
  on 100 Million Labels for Related Searches}. In
  \bibinfo{booktitle}{\emph{Proceedings of the Twelfth ACM International
  Conference on Web Search and Data Mining}} (Melbourne VIC, Australia)
  \emph{(\bibinfo{series}{WSDM ’19})}. \bibinfo{publisher}{Association for
  Computing Machinery}, \bibinfo{pages}{528–536}.
\newblock
\showISBNx{9781450359405}


\bibitem[\protect\citeauthoryear{Jia, Shelhamer, Donahue, Karayev, Long,
  Girshick, Guadarrama, and Darrell}{Jia et~al\mbox{.}}{2014}]%
        {10.1145/2647868.2654889}
\bibfield{author}{\bibinfo{person}{Yangqing Jia}, \bibinfo{person}{Evan
  Shelhamer}, \bibinfo{person}{Jeff Donahue}, \bibinfo{person}{Sergey Karayev},
  \bibinfo{person}{Jonathan Long}, \bibinfo{person}{Ross Girshick},
  \bibinfo{person}{Sergio Guadarrama}, {and} \bibinfo{person}{Trevor Darrell}.}
  \bibinfo{year}{2014}\natexlab{}.
\newblock \showarticletitle{Caffe: Convolutional Architecture for Fast Feature
  Embedding}. In \bibinfo{booktitle}{\emph{Proceedings of the 22nd ACM
  International Conference on Multimedia}} (Orlando, Florida, USA)
  \emph{(\bibinfo{series}{MM ’14})}. \bibinfo{publisher}{Association for
  Computing Machinery}, \bibinfo{pages}{675–678}.
\newblock
\showISBNx{9781450330633}


\bibitem[\protect\citeauthoryear{Jiang, Wu, and Fu}{Jiang
  et~al\mbox{.}}{2016}]%
        {10.1145/2964284.2967182}
\bibfield{author}{\bibinfo{person}{Shuhui Jiang}, \bibinfo{person}{Yue Wu},
  {and} \bibinfo{person}{Yun Fu}.} \bibinfo{year}{2016}\natexlab{}.
\newblock \showarticletitle{Deep Bi-Directional Cross-Triplet Embedding for
  Cross-Domain Clothing Retrieval}. In \bibinfo{booktitle}{\emph{Proceedings of
  the 24th ACM International Conference on Multimedia}} (Amsterdam, The
  Netherlands) \emph{(\bibinfo{series}{MM ’16})}.
  \bibinfo{publisher}{Association for Computing Machinery},
  \bibinfo{pages}{52–56}.
\newblock
\showISBNx{9781450336031}


\bibitem[\protect\citeauthoryear{Kendall, Gal, and Cipolla}{Kendall
  et~al\mbox{.}}{2017}]%
        {DBLP:journals/corr/KendallGC17}
\bibfield{author}{\bibinfo{person}{Alex Kendall}, \bibinfo{person}{Yarin Gal},
  {and} \bibinfo{person}{Roberto Cipolla}.} \bibinfo{year}{2017}\natexlab{}.
\newblock \showarticletitle{Multi-Task Learning Using Uncertainty to Weigh
  Losses for Scene Geometry and Semantics}.
\newblock \bibinfo{journal}{\emph{CoRR}}  \bibinfo{volume}{abs/1705.07115}
  (\bibinfo{year}{2017}).
\newblock
\showeprint[arxiv]{1705.07115}
\urldef\tempurl%
\url{http://arxiv.org/abs/1705.07115}
\showURL{%
\tempurl}


\bibitem[\protect\citeauthoryear{{Li}, {Wang}, {Lyu}, and {Shi}}{{Li}
  et~al\mbox{.}}{2020}]%
        {9051843}
\bibfield{author}{\bibinfo{person}{H. {Li}}, \bibinfo{person}{Y. {Wang}},
  \bibinfo{person}{Z. {Lyu}}, {and} \bibinfo{person}{J. {Shi}}.}
  \bibinfo{year}{2020}\natexlab{}.
\newblock \showarticletitle{Multi-task Learning for Recommendation over
  Heterogeneous Information Network}.
\newblock \bibinfo{journal}{\emph{IEEE Transactions on Knowledge and Data
  Engineering}} (\bibinfo{year}{2020}), \bibinfo{pages}{1--1}.
\newblock


\bibitem[\protect\citeauthoryear{Liu, Ounis, Macdonald, and Meng}{Liu
  et~al\mbox{.}}{2020}]%
        {10.1145/3397271.3401252}
\bibfield{author}{\bibinfo{person}{Siwei Liu}, \bibinfo{person}{Iadh Ounis},
  \bibinfo{person}{Craig Macdonald}, {and} \bibinfo{person}{Zaiqiao Meng}.}
  \bibinfo{year}{2020}\natexlab{}.
\newblock \showarticletitle{A Heterogeneous Graph Neural Model for Cold-Start
  Recommendation}. In \bibinfo{booktitle}{\emph{Proceedings of the 43rd
  International ACM SIGIR Conference on Research and Development in Information
  Retrieval}} (Virtual Event, China) \emph{(\bibinfo{series}{SIGIR ’20})}.
  \bibinfo{publisher}{Association for Computing Machinery},
  \bibinfo{pages}{2029–2032}.
\newblock
\showISBNx{9781450380164}


\bibitem[\protect\citeauthoryear{{Liu}, {Song}, {Liu}, {Xu}, {Lu}, and
  {Yan}}{{Liu} et~al\mbox{.}}{2012}]%
        {6248071}
\bibfield{author}{\bibinfo{person}{S. {Liu}}, \bibinfo{person}{Z. {Song}},
  \bibinfo{person}{G. {Liu}}, \bibinfo{person}{C. {Xu}}, \bibinfo{person}{H.
  {Lu}}, {and} \bibinfo{person}{S. {Yan}}.} \bibinfo{year}{2012}\natexlab{}.
\newblock \showarticletitle{Street-to-shop: Cross-scenario clothing retrieval
  via parts alignment and auxiliary set}. In \bibinfo{booktitle}{\emph{2012
  IEEE Conference on Computer Vision and Pattern Recognition}}.
  \bibinfo{pages}{3330--3337}.
\newblock


\bibitem[\protect\citeauthoryear{McAuley, Targett, Shi, and van~den
  Hengel}{McAuley et~al\mbox{.}}{2015}]%
        {McATarShiHen15}
\bibfield{author}{\bibinfo{person}{Julian McAuley},
  \bibinfo{person}{Christopher Targett}, \bibinfo{person}{Qinfeng Shi}, {and}
  \bibinfo{person}{Anton van~den Hengel}.} \bibinfo{year}{2015}\natexlab{}.
\newblock \showarticletitle{Image-based recommendations on styles and
  substitutes}. In \bibinfo{booktitle}{\emph{SIGIR}}.
\newblock


\bibitem[\protect\citeauthoryear{Mikolov, Chen, Corrado, and Dean}{Mikolov
  et~al\mbox{.}}{2013a}]%
        {DBLP:journals/corr/abs-1301-3781}
\bibfield{author}{\bibinfo{person}{Tomas Mikolov}, \bibinfo{person}{Kai Chen},
  \bibinfo{person}{Greg Corrado}, {and} \bibinfo{person}{Jeffrey Dean}.}
  \bibinfo{year}{2013}\natexlab{a}.
\newblock \showarticletitle{Efficient Estimation of Word Representations in
  Vector Space}.
\newblock \bibinfo{journal}{\emph{CoRR}}  \bibinfo{volume}{abs/1301.3781}
  (\bibinfo{year}{2013}).
\newblock
\showeprint[arxiv]{1301.3781}


\bibitem[\protect\citeauthoryear{Mikolov, Sutskever, Chen, Corrado, and
  Dean}{Mikolov et~al\mbox{.}}{2013b}]%
        {Mikolov:2013:DRW:2999792.2999959}
\bibfield{author}{\bibinfo{person}{Tomas Mikolov}, \bibinfo{person}{Ilya
  Sutskever}, \bibinfo{person}{Kai Chen}, \bibinfo{person}{Greg Corrado}, {and}
  \bibinfo{person}{Jeffrey Dean}.} \bibinfo{year}{2013}\natexlab{b}.
\newblock \showarticletitle{Distributed Representations of Words and Phrases
  and Their Compositionality}. In \bibinfo{booktitle}{\emph{Proceedings of the
  26th International Conference on Neural Information Processing Systems -
  Volume 2}} (Lake Tahoe, Nevada) \emph{(\bibinfo{series}{NIPS'13})}.
  \bibinfo{publisher}{Curran Associates Inc.}, \bibinfo{address}{USA},
  \bibinfo{pages}{3111--3119}.
\newblock


\bibitem[\protect\citeauthoryear{Oord, Li, and Vinyals}{Oord
  et~al\mbox{.}}{2018}]%
        {oord2018representation}
\bibfield{author}{\bibinfo{person}{Aaron van~den Oord}, \bibinfo{person}{Yazhe
  Li}, {and} \bibinfo{person}{Oriol Vinyals}.} \bibinfo{year}{2018}\natexlab{}.
\newblock \showarticletitle{Representation learning with contrastive predictive
  coding}.
\newblock \bibinfo{journal}{\emph{arXiv preprint arXiv:1807.03748}}
  (\bibinfo{year}{2018}).
\newblock


\bibitem[\protect\citeauthoryear{Perozzi, Al-Rfou, and Skiena}{Perozzi
  et~al\mbox{.}}{2014}]%
        {perozzi2014deepwalk}
\bibfield{author}{\bibinfo{person}{Bryan Perozzi}, \bibinfo{person}{Rami
  Al-Rfou}, {and} \bibinfo{person}{Steven Skiena}.}
  \bibinfo{year}{2014}\natexlab{}.
\newblock \showarticletitle{Deepwalk: Online learning of social
  representations}. In \bibinfo{booktitle}{\emph{Proceedings of the 20th ACM
  SIGKDD international conference on Knowledge discovery and data mining}}.
  \bibinfo{pages}{701--710}.
\newblock


\bibitem[\protect\citeauthoryear{Rawat and Kankanhalli}{Rawat and
  Kankanhalli}{2016}]%
        {rawat2016contagnet}
\bibfield{author}{\bibinfo{person}{Yogesh~Singh Rawat} {and}
  \bibinfo{person}{Mohan~S Kankanhalli}.} \bibinfo{year}{2016}\natexlab{}.
\newblock \showarticletitle{ConTagNet: Exploiting user context for image tag
  recommendation}. In \bibinfo{booktitle}{\emph{Proceedings of the 24th ACM
  international conference on Multimedia}}. \bibinfo{pages}{1102--1106}.
\newblock


\bibitem[\protect\citeauthoryear{Rendle, Freudenthaler, Gantner, and
  Schmidt{-}Thieme}{Rendle et~al\mbox{.}}{2012}]%
        {DBLP:journals/corr/abs-1205-2618}
\bibfield{author}{\bibinfo{person}{Steffen Rendle}, \bibinfo{person}{Christoph
  Freudenthaler}, \bibinfo{person}{Zeno Gantner}, {and} \bibinfo{person}{Lars
  Schmidt{-}Thieme}.} \bibinfo{year}{2012}\natexlab{}.
\newblock \showarticletitle{{BPR:} Bayesian Personalized Ranking from Implicit
  Feedback}.
\newblock \bibinfo{journal}{\emph{CoRR}}  \bibinfo{volume}{abs/1205.2618}
  (\bibinfo{year}{2012}).
\newblock
\showeprint[arxiv]{1205.2618}
\urldef\tempurl%
\url{http://arxiv.org/abs/1205.2618}
\showURL{%
\tempurl}


\bibitem[\protect\citeauthoryear{Saveski and Mantrach}{Saveski and
  Mantrach}{2014}]%
        {10.1145/2645710.2645751}
\bibfield{author}{\bibinfo{person}{Martin Saveski} {and} \bibinfo{person}{Amin
  Mantrach}.} \bibinfo{year}{2014}\natexlab{}.
\newblock \showarticletitle{Item Cold-Start Recommendations: Learning Local
  Collective Embeddings}. In \bibinfo{booktitle}{\emph{Proceedings of the 8th
  ACM Conference on Recommender Systems}} (Foster City, Silicon Valley,
  California, USA) \emph{(\bibinfo{series}{RecSys ’14})}.
  \bibinfo{publisher}{Association for Computing Machinery},
  \bibinfo{pages}{89–96}.
\newblock
\showISBNx{9781450326681}


\bibitem[\protect\citeauthoryear{Schein, Popescul, Ungar, and Pennock}{Schein
  et~al\mbox{.}}{2002}]%
        {10.1145/564376.564421}
\bibfield{author}{\bibinfo{person}{Andrew~I. Schein},
  \bibinfo{person}{Alexandrin Popescul}, \bibinfo{person}{Lyle~H. Ungar}, {and}
  \bibinfo{person}{David~M. Pennock}.} \bibinfo{year}{2002}\natexlab{}.
\newblock \showarticletitle{Methods and Metrics for Cold-Start
  Recommendations}. In \bibinfo{booktitle}{\emph{Proceedings of the 25th Annual
  International ACM SIGIR Conference on Research and Development in Information
  Retrieval}} (Tampere, Finland) \emph{(\bibinfo{series}{SIGIR ’02})}.
  \bibinfo{publisher}{Association for Computing Machinery},
  \bibinfo{pages}{253–260}.
\newblock
\showISBNx{1581135610}


\bibitem[\protect\citeauthoryear{Shankar, Narumanchi, Ananya, Kompalli, and
  Chaudhury}{Shankar et~al\mbox{.}}{2017}]%
        {DBLP:journals/corr/ShankarNAKC17}
\bibfield{author}{\bibinfo{person}{Devashish Shankar}, \bibinfo{person}{Sujay
  Narumanchi}, \bibinfo{person}{H.~A. Ananya}, \bibinfo{person}{Pramod
  Kompalli}, {and} \bibinfo{person}{Krishnendu Chaudhury}.}
  \bibinfo{year}{2017}\natexlab{}.
\newblock \showarticletitle{Deep Learning based Large Scale Visual
  Recommendation and Search for E-Commerce}.
\newblock \bibinfo{journal}{\emph{CoRR}}  \bibinfo{volume}{abs/1703.02344}
  (\bibinfo{year}{2017}).
\newblock
\showeprint[arxiv]{1703.02344}
\urldef\tempurl%
\url{http://arxiv.org/abs/1703.02344}
\showURL{%
\tempurl}


\bibitem[\protect\citeauthoryear{{Shi}, {Hu}, {Zhao}, and {Yu}}{{Shi}
  et~al\mbox{.}}{2019}]%
        {8355676}
\bibfield{author}{\bibinfo{person}{C. {Shi}}, \bibinfo{person}{B. {Hu}},
  \bibinfo{person}{W.~X. {Zhao}}, {and} \bibinfo{person}{P.~S. {Yu}}.}
  \bibinfo{year}{2019}\natexlab{}.
\newblock \showarticletitle{Heterogeneous Information Network Embedding for
  Recommendation}.
\newblock \bibinfo{journal}{\emph{IEEE Transactions on Knowledge and Data
  Engineering}} \bibinfo{volume}{31}, \bibinfo{number}{2}
  (\bibinfo{year}{2019}), \bibinfo{pages}{357--370}.
\newblock


\bibitem[\protect\citeauthoryear{Vasile, Smirnova, and Conneau}{Vasile
  et~al\mbox{.}}{2016}]%
        {10.1145/2959100.2959160}
\bibfield{author}{\bibinfo{person}{Flavian Vasile}, \bibinfo{person}{Elena
  Smirnova}, {and} \bibinfo{person}{Alexis Conneau}.}
  \bibinfo{year}{2016}\natexlab{}.
\newblock \showarticletitle{Meta-Prod2Vec: Product Embeddings Using
  Side-Information for Recommendation}. In
  \bibinfo{booktitle}{\emph{Proceedings of the 10th ACM Conference on
  Recommender Systems}} (Boston, Massachusetts, USA)
  \emph{(\bibinfo{series}{RecSys ’16})}. \bibinfo{publisher}{Association for
  Computing Machinery}, \bibinfo{pages}{225–232}.
\newblock
\showISBNx{9781450340359}


\bibitem[\protect\citeauthoryear{Wang, Wang, Tang, Shu, Ranganath, and
  Liu}{Wang et~al\mbox{.}}{2017}]%
        {wang2017your}
\bibfield{author}{\bibinfo{person}{Suhang Wang}, \bibinfo{person}{Yilin Wang},
  \bibinfo{person}{Jiliang Tang}, \bibinfo{person}{Kai Shu},
  \bibinfo{person}{Suhas Ranganath}, {and} \bibinfo{person}{Huan Liu}.}
  \bibinfo{year}{2017}\natexlab{}.
\newblock \showarticletitle{What your images reveal: Exploiting visual contents
  for point-of-interest recommendation}. In
  \bibinfo{booktitle}{\emph{Proceedings of the 26th International Conference on
  World Wide Web}}. \bibinfo{pages}{391--400}.
\newblock


\bibitem[\protect\citeauthoryear{{Xu}, {Lu}, {Song}, {Yang}, {Shen}, and
  {Li}}{{Xu} et~al\mbox{.}}{2020}]%
        {8771379}
\bibfield{author}{\bibinfo{person}{X. {Xu}}, \bibinfo{person}{H. {Lu}},
  \bibinfo{person}{J. {Song}}, \bibinfo{person}{Y. {Yang}},
  \bibinfo{person}{H.~T. {Shen}}, {and} \bibinfo{person}{X. {Li}}.}
  \bibinfo{year}{2020}\natexlab{}.
\newblock \showarticletitle{Ternary Adversarial Networks With Self-Supervision
  for Zero-Shot Cross-Modal Retrieval}.
\newblock \bibinfo{journal}{\emph{IEEE Transactions on Cybernetics}}
  \bibinfo{volume}{50} (\bibinfo{year}{2020}), \bibinfo{pages}{2400--2413}.
\newblock


\bibitem[\protect\citeauthoryear{Ying, He, Chen, Eksombatchai, Hamilton, and
  Leskovec}{Ying et~al\mbox{.}}{2018}]%
        {10.1145/3219819.3219890}
\bibfield{author}{\bibinfo{person}{Rex Ying}, \bibinfo{person}{Ruining He},
  \bibinfo{person}{Kaifeng Chen}, \bibinfo{person}{Pong Eksombatchai},
  \bibinfo{person}{William~L. Hamilton}, {and} \bibinfo{person}{Jure
  Leskovec}.} \bibinfo{year}{2018}\natexlab{}.
\newblock \showarticletitle{Graph Convolutional Neural Networks for Web-Scale
  Recommender Systems}. In \bibinfo{booktitle}{\emph{Proceedings of the 24th
  ACM SIGKDD International Conference on Knowledge Discovery and Data Mining}}
  (London, United Kingdom) \emph{(\bibinfo{series}{KDD ’18})}.
  \bibinfo{publisher}{Association for Computing Machinery},
  \bibinfo{pages}{974–983}.
\newblock
\showISBNx{9781450355520}


\bibitem[\protect\citeauthoryear{Zhang, Yao, Sun, and Tay}{Zhang
  et~al\mbox{.}}{2019}]%
        {10.1145/3285029}
\bibfield{author}{\bibinfo{person}{Shuai Zhang}, \bibinfo{person}{Lina Yao},
  \bibinfo{person}{Aixin Sun}, {and} \bibinfo{person}{Yi Tay}.}
  \bibinfo{year}{2019}\natexlab{}.
\newblock \showarticletitle{Deep Learning Based Recommender System: A Survey
  and New Perspectives}.
\newblock \bibinfo{journal}{\emph{ACM Comput. Surv.}} \bibinfo{volume}{52},
  \bibinfo{number}{1}, Article \bibinfo{articleno}{5} (\bibinfo{date}{Feb.}
  \bibinfo{year}{2019}).
\newblock
\showISSN{0360-0300}


\end{thebibliography}

\appendix

\end{document}